\documentclass[aps,twocolumn,showpacs,floatfix]{revtex4}
\usepackage{graphicx}
\usepackage{dcolumn}
\usepackage{bm}

\begin{document}
\title{Anomalous temperature-induced volume contraction in GeTe}
\author{Tapan Chatterji$^1$, C. M. N. Kumar$^{2,3}$ and Urszula D. Wdowik$^4$ }
\address{$^1$Institut Laue-Langevin, B.P. 156, 38042 Grenoble Cedex 9, France\\
$^2$ J\"ulich Centre for Neutron Science, Forschungszentrum J\"ulich, D-52425 J\"ulich, Germany\\
$^3$Neutron Sciences Directorate, Spallation Neutron Source, Oak Ridge National Laboratory, Oak Ridge, TN 37831-6475, USA \\
$^4$Institute of Technology, Pedagogical University, Podchorazych 2, PL-30084 Krak\'ow, Poland }
\begin{abstract}
The recent surge of interest in phase change materials GeTe, Ge$_2$Sb$_2$Te$_5$, and related compounds motivated us to revisit the structural phase transition in GeTe in more details than was done before. Rhombohedral-to-cubic ferroelectric phase transition in GeTe has been studied by high resolution neutron powder diffraction on a spallation neutron source. We determined the temperature dependence of the structural parameters in a wide temperature range extending from 309 to 973 K. Results of our studies clearly show an anomalous volume contraction of 0.6\% at the phase transition from the rhombohedral to cubic phase. In order to better understand the phase transition and the associated anomalous volume decrease in GeTe we have performed phonon calculations based on the density functional theory. Results of the present investigations are also discussed with respect to the experimental data obtained for single crystals of GeTe.

\end{abstract}
\pacs{} 
\maketitle

\section{Introduction}
Phase change materials possess unique properties that hold considerable promise for applications in data storage \cite{wuttig07}. They can be rapidly and reversibly switched between amorphous and crystalline states, which differ substantially in their properties. Recently materials such as Ge$_2$Sb$_2$Te$_5$ and Ag and In doped Sb$_2$Te have been discovered to crystallize rapidly enough to enable competitive solutions for rewritable optical data storage. 

The related binary material GeTe has drawn considerable interest because of its higher crystalline temperatures and better data retention at high temperatures compared to Ge$_2$Sb$_2$Te$_5$. Apart from the application in data storage, GeTe may have potential use as a thermoelectric material \cite{levin13}. These have motivated us to revisit the high temperature ferroelectric phase transition in GeTe.  
    
GeTe is a narrow band-gap semiconductor \cite{tsu68} and is ferroelectric at room temperature with a Curie temperature of  about 705~K. The low temperature ferroelectric phase has a rhombohedraly distorted NaCl type crystal structure  with the space group $R3m$ \cite {schubert51,schubert51a,goldak66,steigmeier70,chattopadhyay87}. Structural distortions involve relative displacement of the Ge and Te sublattices along the body cell diagonal and subsequent rhombohedral shear deformation along $[111]$ direction which changes the rhombohedral angle from its fcc value of $60^{\circ}$ to $\alpha$. The Ge and Te atoms are six fold coordinated by each other with three shorter (2.83~\AA) and three longer (3.15~\AA) bonds. This is often described as Peierls distortion \cite{peierls55} due to reduced coupling between the $p$-type orbitals that constitute basis of the bonding in GeTe. 

GeTe undergoes a ferroelectric phase transition in which the low temperature rhombohedral $R3m$ structure transform to the cubic $Fm\bar{3}m$ structure at high temperature of about 600--700 K  \cite{steigmeier70,chattopadhyay87}. The transition temperature $T_c$ depends on the sample stoichiometry and carrier concentration \cite{chattopadhyay87}. Ferroelectric phase transition in GeTe was considered to be  displacive in its origin \cite{chattopadhyay87,steigmeier70,rabe87,wdowik14}. Recently, however, the displacive character of the rhombohedral-to-cubic phase transition in GeTe has been contested by Fons  \textit{at al.} \cite{fons10} and Matsunaga \textit{et al.} \cite{matsunaga11}. According to their studies the displacive nature of this phase transition was due to the misinterpretation of the Bragg diffraction results as the structure determination based only on the Bragg intensities gives information about the average structure, but not about the system local structure. Information about the local structure can be obtained from the total scattering data including the diffuse scattering up to a very high $Q$ value and from the pair-distribution function (PDF) analysis \cite{egami03}. Such investigations has been performed on the X-ray diffraction data by Matsunaga \textit{et al.} \cite{matsunaga11} and lead to the conclusion that in the local scale the high temperature phase of GeTe still exhibits distinct short and long Ge--Te bonds, contrary to the conventional structure refinement with only the Bragg intensities considered which suggests that the high temperature cubic phase reveals solely one type of the Ge--Te bond. Additionally, the two distinct bond distances observed in the local scale hardly change across the rhombohedral-to-cubic phase transition. Results of Matsunaga \textit{et al.} \cite{matsunaga11} were also supported by the EXAFS studies of Fons \textit{et al.} \cite{fons10}. These findings suggested that the phase transition in GeTe is not displacive but order-disorder type. On the other hand, the most recent lattice dynamical calculations \cite{wdowik14} based on density functional theory (DFT) show that the rhombohedral-to-cubic phase transition in GeTe is indeed displacive in its origin and becomes driven by the condensation of exactly three components of the triply degenerate optical transverse soft phonon mode at the Brillouin zone center. Moreover, the displacive character of the phase transition in GeTe has been further supported by the recent electron and X-ray diffraction studies as well as the Raman scattering experiments of Polking \textit{et al.} \cite{polking10}. 
   
We have revisited the ferroelectric phase transition in GeTe by the high resolution neutron powder diffraction on a modern high power spallation neutron source and determined the temperature variation of the lattice parameters, unit cell volume, positional parameters, and bond distances more accurately and in much finer temperature steps across the rhombohedral-to-cubic phase transition than were done for this compound before \cite{chattopadhyay87}. Also, the DFT phonon calculations were performed to better understand the phase transition and the associated anomalous volume contraction in GeTe.
      
\section{Experimental and Calculation Methods}   
GeTe powder samples were obtained from Alfa Aesar. The samples are claimed to be 99.999\% pure and are 200 Mesh. We checked the samples by the X-ray powder diffraction and found the presence of small amount of GeO$_2$ impurity. Neutron powder diffraction (NPD) measurements were performed on the time-of-flight powder diffractometer, POWGEN, located at the Spallation Neutron Source at Oak Ridge National Laboratory. The data were collected with neutrons of central wavelength 1.333 {\AA}, covering $d$-spacing range from 0.42 to 5.4 {\AA}. Approximately 5~g of GeTe sample was loaded in a vanadium container of 10 mm diameter and measured in a traditional ILL furnace within the temperature range of 309--973~K. Structure refinement was carried out using the \textsc{FullProf} suite \cite{rodriguez10}.

The present theoretical studies use the DFT method implemented in the \textsc{vasp} code \cite{kresse99} together with the direct method approach \cite{parlinski97} to provide necessary input data to calculate the temperature dependence of the mean-squared vibrational amplitudes ($U_{ij}$) of the Ge and Te atoms in the low and high temperature phases of GeTe within the harmonic theory. The $U_{ij}$ tensor is obtained from the calculated diagonal and off-diagonal partial phonon densities of states \cite{wdowik10}. The volume thermal expansions for the low and high temperature structures of GeTe are evaluated according the quasiharmonic approximation (QHA) \cite{wallace72}. Details of the lattice dynamics calculations on GeTe system can be found in Ref.~\cite{wdowik14}. The pair-distribution functions of GeTe were obtained using the \textsc{PDFgui} program \cite{Pdfgui}.
 
\section{Experimental Results}
\subsection{Crystal structure and unit cell transformation}
\begin{figure}
\includegraphics[width=0.5\columnwidth]{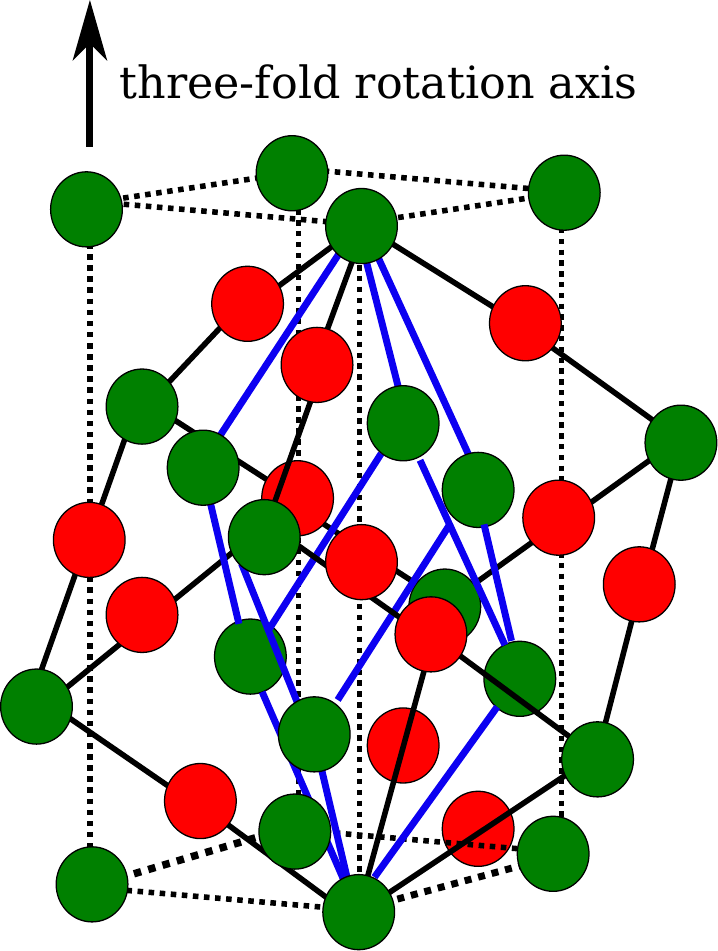}
\caption {(Color online) Structure of the low temperature phase of GeTe (space group $R3m$) shown in the pseudo-cubic (thick black lines), hexagonal (dotted lines), and rhombohedral (thick blue/gray lines) representations.}
 \label{structure}
\end{figure}

Before we discuss the present NPD results and compare them with the results of neutron diffraction experiments on single crystals \cite{chattopadhyay87}, we describe the different units cells used in the two investigations as well as the relationship between them. The low temperature structure of GeTe with the space group $R3m$ can be expressed in the pseudo-cubic, rhombohedral or hexagonal crystallographic representations as illustrated in Fig.~\ref{structure}. The distorted rocksalt structure with the lattice parameter $a_c$ and angle $\alpha$ is related to hexagonal unit cell with the lattice constants $a$ and $c$ via the relation $a = 2a_c\sin (\alpha/2)$ and $c=a_c \sqrt{3+6 \cos \alpha}$  \cite{okeefee96}. In the hexagonal representation the Ge and Te atoms occupy $3a(0,0,x)$ and $3a(0,0,1-x)$ Wyckoff positions, respectively.  In addition, the distortion parameter $\Delta x = 0.25 - x_{\mathrm{Ge}}=x_{\mathrm{Te}} - 0.75$ describes the relative shift of the Ge and Te sublattices from the values $x_\mathrm{Ge}=0.25$ and $x_\mathrm{Te}=0.75$ which are characteristic for the cubic GeTe structure. The degree of distortion from the cubic NaCl-type structure is also reflected in the deviation  $\Delta \alpha$ of the rhombohedral angle $\alpha$ from the cubic value of 90$^{\circ}$ ($\Delta \alpha = 90^{\circ} - \alpha$).

We recall that Chattopadhyay \textit{et al.} \cite{chattopadhyay87} performed structural refinement of their single crystal data using the pseudo-cubic cell with the $F13m$ setting whereas we have done structure refinement of our NPD data in the hexagonal setting of a rhombohedral cell. At 309~K, the hexagonal cell is described by the lattice parameters $a_h = 4.1651$ {\AA} and $c_h = 10.6704$ {\AA}, corresponding to the pseudo-cubic cell with the lattice parameters $a_c = 5.9818$ {\AA} and $\alpha_c = 88.2615 ^{\circ}$. Although we refined the powder diffraction data in the hexagonal cell we choose the pseudo-cubic representation as the most convenient to make comparison between the present results and those of Chattopadhyay \textit{et al.} \cite{chattopadhyay87}. The usage of the pseudo-cubic representations is also advantageous because $\alpha=90^{\circ}$ in the high temperature phase.

\subsection{Neutron Powder Diffraction Results}

\begin{figure}
\includegraphics[width=0.99\columnwidth]{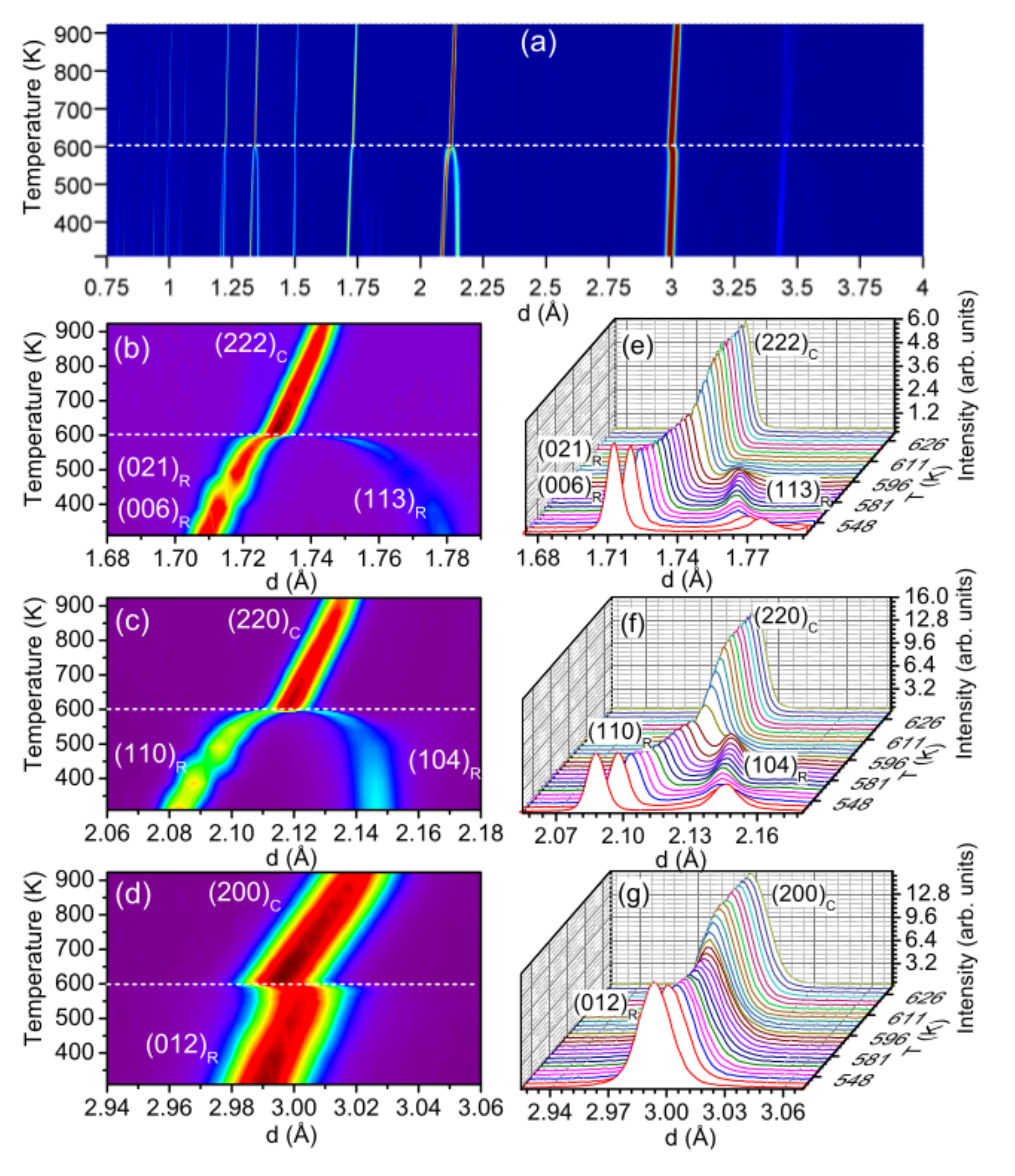}
\caption {(Color online) Temperature dependence of diffraction diagrams for GeTe measured at wide $d$-range (a) and $d$-ranges corresponding respectively to the cubic $(222)_c$ (b), $(220)_c$ (c), and $(200)_c$ (d) reflections. Note the phase transition at $T_c = 600$~K, development of the rhombohedral distortions below $T_c$, and vanishing splitting of the diffraction peaks at the onset of the rhombohedral-to-cubic phase transition.}
 \label{diagram}
\end{figure}

Figures \ref{diagram} (a)--(d) show the temperature variation of the diffraction diagram represented by contour plots. The large $d$-range is presented in Fig.~\ref{diagram}(a), while the $d$-ranges corresponding to the cubic $(222)_c$, $(220)_c$, and $(200)_c$ reflections are depicted in Fig.~\ref{diagram}(c)-(d). These reflections are split due to distortions of the Peierls type below $T_c = 600$~K. The splitting progressively diminishes with increasing temperature and finally the double-peak structure disappears at the onset of structural transformation. The single peak observed above 600~K is a direct evidence of the cubic symmetry of GeTe.   

\begin{figure}
\includegraphics[width=0.99\columnwidth]{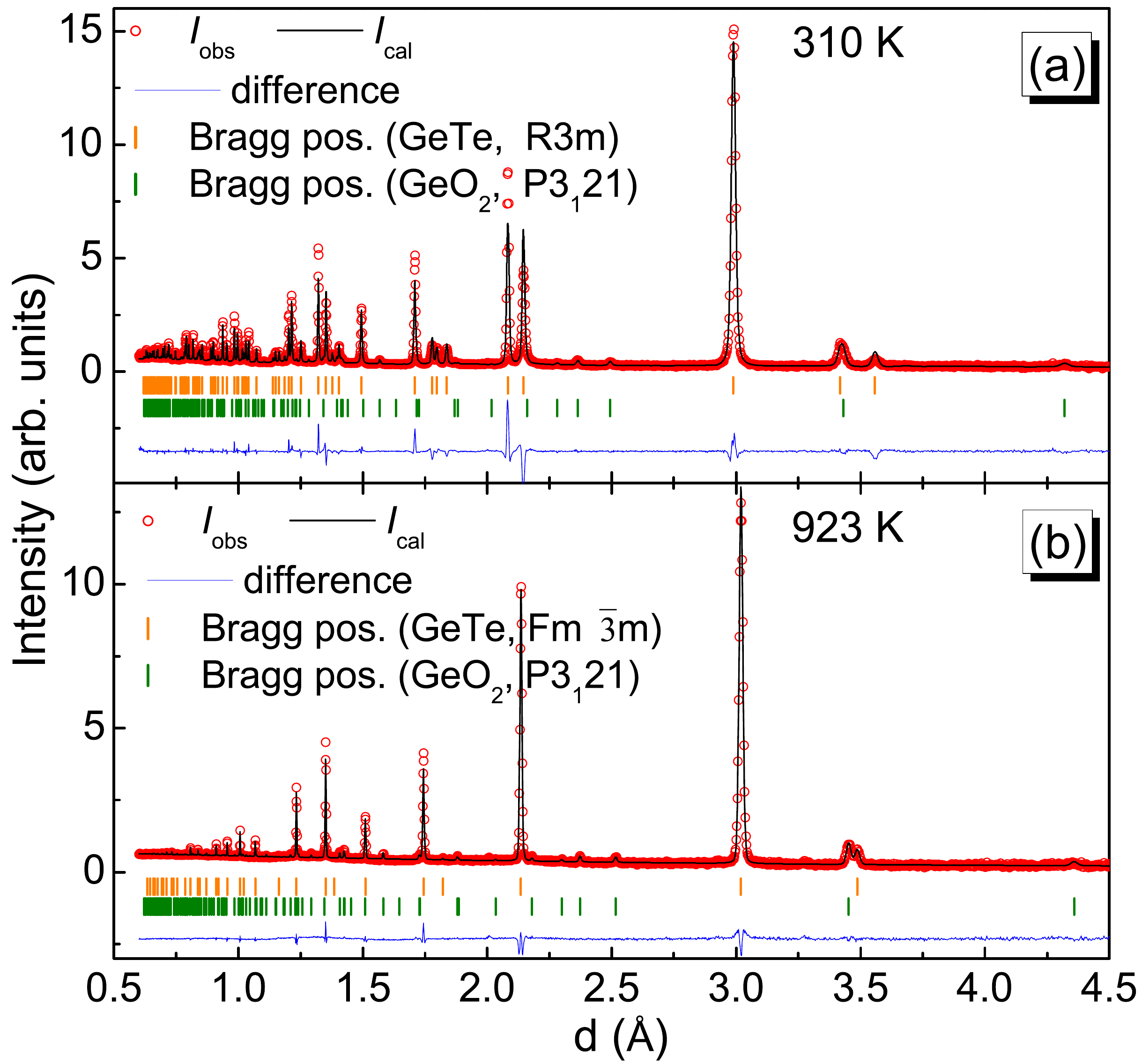}
 \caption {(Color online) Results of the Rietveld profile refinement for the low temperature rhombohedral phase of GeTe (a) and the high temperature cubic phase of GeTe (b). The refinements were performed by taking into account the impurity phase GeO$_2$. The orange and green small vertical lines indicate the positions of the Bragg peaks corresponding to the GeTe and GeO$_2$ phases, respectively. The blue curves at the bottom denote differences between the measured and calculated intensities. The conventional Rietveld discrepancy parameters for the refinement (a) were $R_p = 20.9\%, R_{wp} = 21.3\%, R_e = 8.48\%$ and $\chi^2= 6.323$ and those for the refinement (b) were  $R_p = 26.6\%, R_{wp} = 18.9\%, R_e = 11.9\%$ and $\chi^2= 2.509$.}
\label{Rietveld}
\end{figure}

Diffraction intensities of the GeTe samples have been refined together with the impurity GeO$_2$ phase. Results of the Rietveld refinements performed at 310~K (low temperature structure) and 923~K (high temperature structure) are shown in Fig.~\ref{Rietveld}. The conventional Rietveld discrepancy parameters for the refinement (a) were $R_p = 20.9\%, R_{wp} = 21.3\%, R_e = 8.48\%$ and $\chi^2= 6.323$ and those for the refinement (b) were  $R_p = 26.6\%, R_{wp} = 18.9\%, R_e = 11.9\%$ and $\chi^2= 2.509$. The suffixes $p$, $wp$ and $e$ mean profile, weighted profile and expected from the counting statistics, respectively. We preferred to call these parameters discrepancy rather than agreement parameters because these parameters are larger when the discrepancy (and not the agreement) is larger. It occurs that the impurity phase does not affect the refined parameters neither of the low nor high temperature GeTe phases. We performed such refinements for the data measured at several temperatures from $\sim$300~K to $\sim$1000 K. The parameters obtained from these refinements will be described and discussed in the following section.

\section{discussion}
\begin{figure}
\includegraphics[width=0.99\columnwidth]{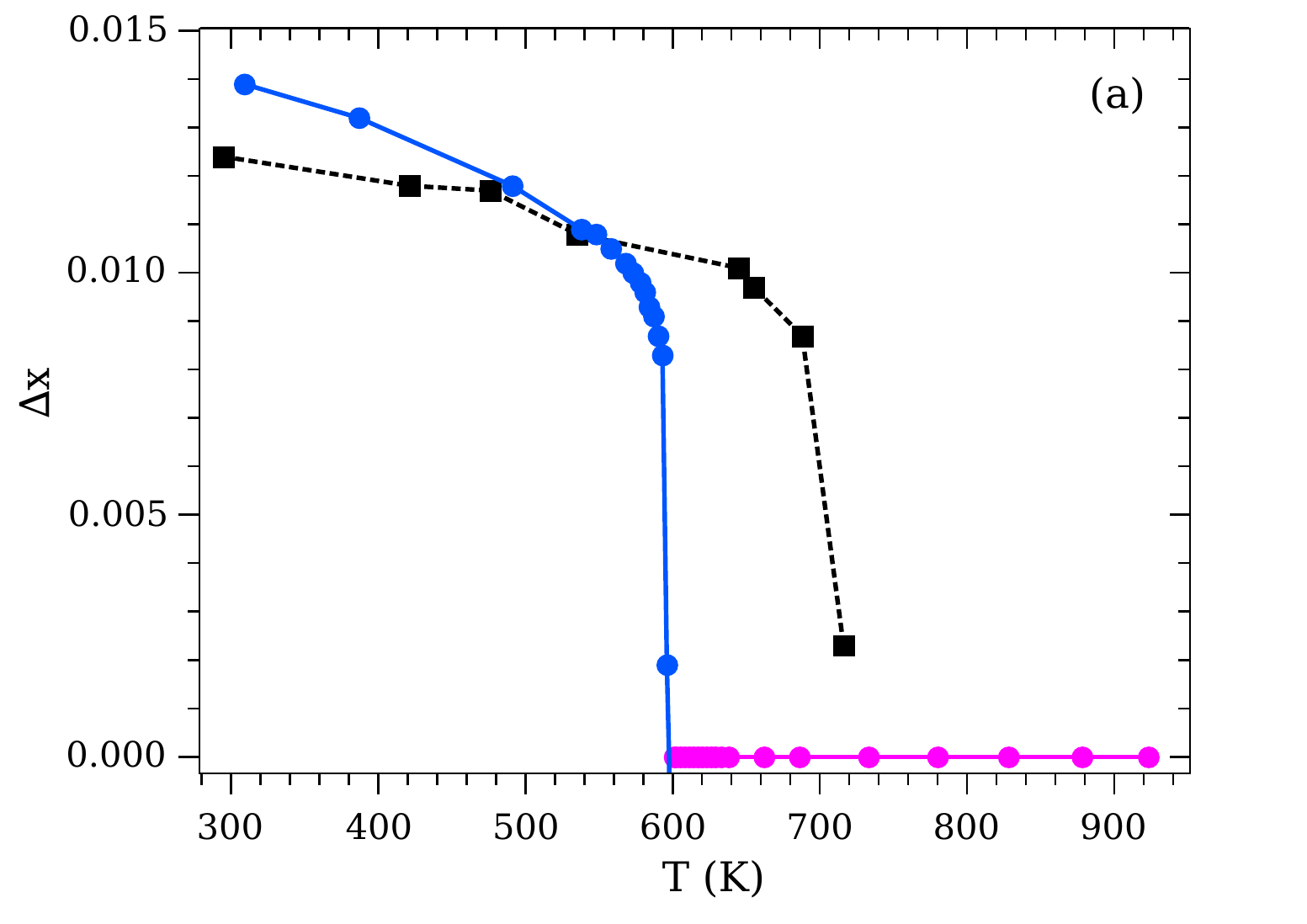}
\includegraphics[width=0.99\columnwidth]{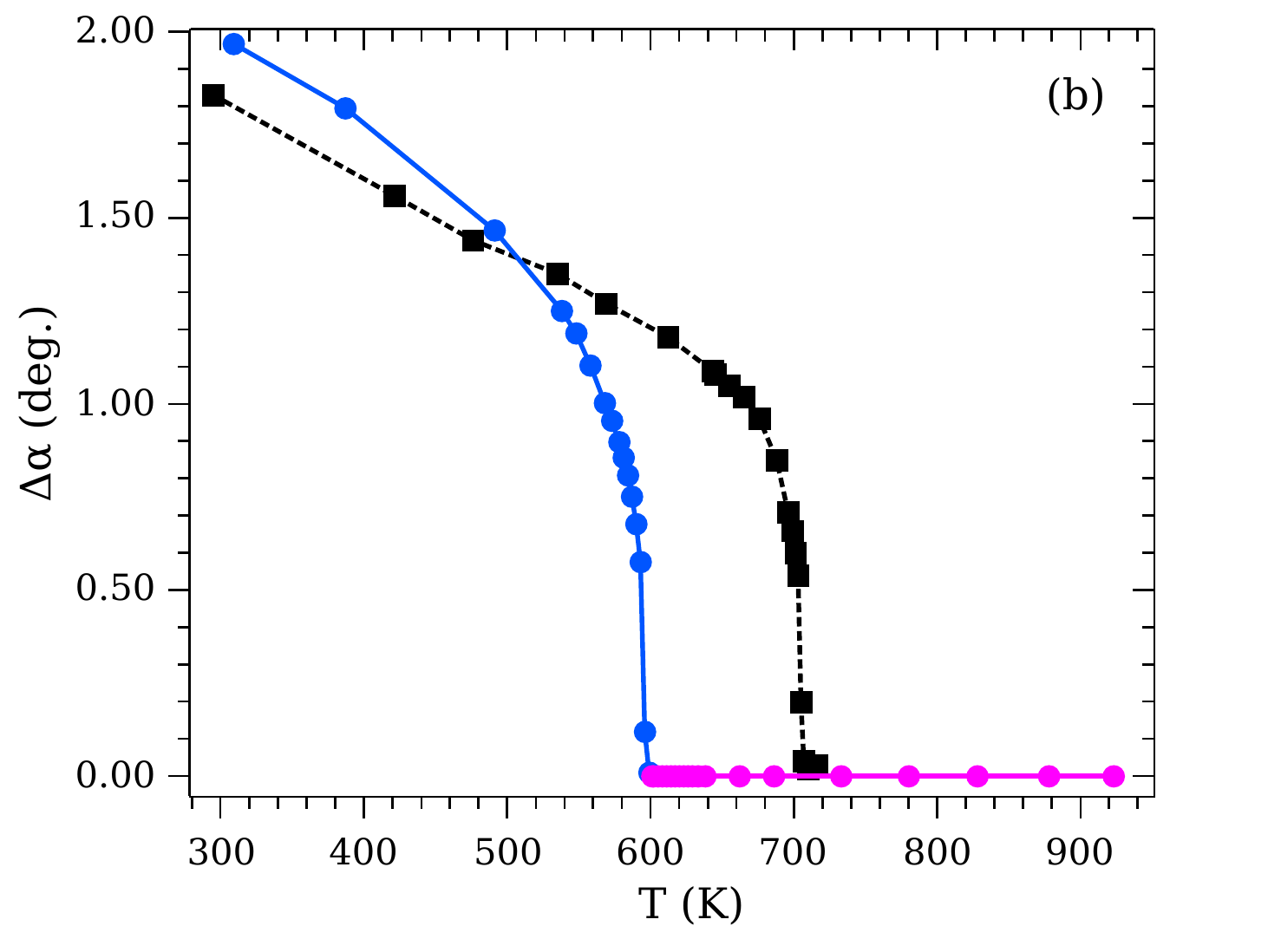}
\caption {(Color online) Temperature variations of (a) distortion parameter $\Delta x$ and (b) deviation $\Delta \alpha$ of the distortion angle from $90^{\circ}$.  The present NPD data (circles) are compared to the neutron diffraction experiments performed on single crystals by Chattopadhyay \textit{et al.} \cite{chattopadhyay87} (squares).} The continuous lines are just guides to the eye.
 \label{distortions}
\end{figure}

The rhombohedral-to-cubic phase transition in GeTe is characterized by pronounced changes in the positional parameters $x_{\mathrm{Ge}}$ and  $x_{\mathrm{Te}}$. The $x_{\mathrm{Ge}}$ ($x_{\mathrm{Te}}$) increases (decreases) with increasing temperature. The initial increase (decrease) in $x_{\mathrm{Ge}}$ ($x_{\mathrm{Te}}$) is almost linear up to $\sim500$~K and becomes non-linear while approaching the transition temperature $T_c=600$~K. One observes sudden raise (drop) of $x_{\mathrm{Ge}}$ ($x_{\mathrm{Te}}$) at $T_c$. Above $T_c$, the $x_{\mathrm{Ge}}$ and $x_{\mathrm{Te}}$ take on the values characteristic for the cubic GeTe phase. These changes are also revealed by $\Delta x$ and $\Delta \alpha$, see Fig.~\ref{distortions}, as the distortion parameters are directly related to $x_{\mathrm{Ge}}$ and $x_{\mathrm{Te}}$. Both $\Delta x$  and $\Delta \alpha$ decrease continuously from $\Delta x = 0.14$ and $\Delta \alpha = 1.98^{\circ}$ at 300~K to zero at $T_c = 600$~K. The $\Delta x$  and $\Delta \alpha$ obtained for single crystals of GeTe \cite{chattopadhyay87} show very similar behavior as that observed for powder samples. They also progressively diminish with temperature, however with smaller slopes than those determined for our powder samples, and finally they approach zero values at 700~K. The $\Delta x = 0$ and $\Delta \alpha = 0$ indicate that the $R3m$ structure undergoes  transformation into the $Fm\bar{3}m$ structure at $T_c$. Here we note the difference in the transition temperature $T_c$ between our powder samples and the single crystal GeTe samples investigated by Chattopadhyay \textit{et al.} \cite{chattopadhyay87}. Such a significant spread in transition temperature of GeTe compound has been known and is usually attributed to the sample stoichiometry as well as the free charge carrier concentration  \cite{tong10}.  

 Distortions $\Delta x$ and $\Delta \alpha$ has been considered respectively as the primary ($Q$) and secondary  ($\epsilon$) order parameters of the phase transition in GeTe \cite{chattopadhyay87}. The continuous shift of the Ge and Te atomic positions $\Delta x$ which is directly related to the amplitude of the soft-phonon mode $\Gamma^-_4$ ($Q$) \cite{wdowik14} breaks the symmetry elements corresponding to the four-fold rotation axis and a mirror plane perpendicular to this axis. It also destroys the symmetry inversion center of the high temperature cubic phase and induces polarization along the three-fold rotation axis. Thus, the primary order parameter $\Delta x$ ($Q$) is the driving force of ferroelectricity appearing in GeTe below $T_c$. On the other hand, the  $\Delta \alpha$ is actually the lattice strain $\epsilon$ which breaks only the symmetry elements corresponding to the four-fold rotation axis and a mirror plane perpendicular to it, but it does not affect the symmetry inversion center of the GeTe system. The neutron diffraction experiments on both powder and single crystal samples show the linear coupling between  $\Delta \alpha$ and $\Delta x^2$, which conforms to the Landau theory \cite{cowley80}. The single crystal data \cite{chattopadhyay87} display, however, slightly smaller slope ($0.89\times10^{-4}$~deg$^{-1}$) than the present NPD data for powder samples ($1.0\times10^{-4}$~deg$^{-1}$). On the other hand, our theoretical calculations give the linear coupling between $\Delta \alpha$ and $\Delta x^2$ of $1.2\times10^{-4}$~deg$^{-1}$.

\begin{figure}[h!]
\includegraphics[width=0.99\columnwidth]{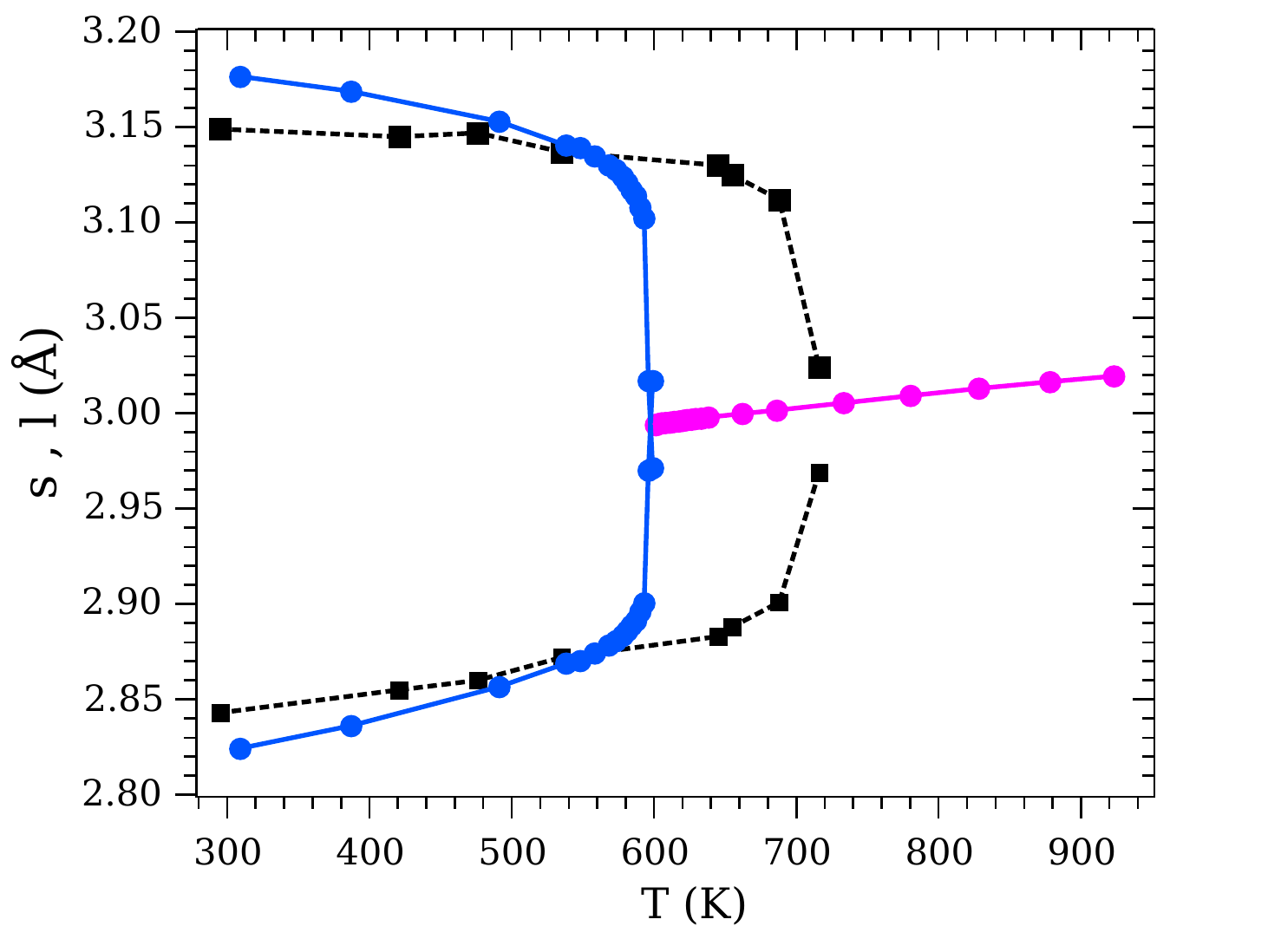}
\caption {(Color online) Temperature dependences of short ($s$) and long ($l$) Ge--Te bonds determined in the present NPD measurements (circles) and those provided by the single crystal neutron diffraction \cite{chattopadhyay87} (squares). The continuous lines are just guides to the eye.}
\label{bondlengths}
\end{figure}

The short ($s$) and long ($l$) bond lengths determined from our neutron measurements on powder samples amount to $s=2.82$~\AA\ and  $l=3.18$~\AA\ at room temperature. They are very close to those obtained from the neutron diffraction studies on single crystal samples \cite{chattopadhyay87}. The $s$ and $l$ bonds in the low temperature rhombohedral phase follow the course of $x_{\mathrm{Ge}}$ and  $x_{\mathrm{Te}}$. They vary smoothly with temperature to reach the average value of $\sim3$~\AA\ above $T_c=600$~K, as shown in Fig.~\ref{bondlengths}. This unique distance is certainly the Ge--Te bond length in the cubic phase which subsequently slightly grows at still higher temperatures due to the thermal expansion of the cubic GeTe lattice. Nevertheless the visible shift between the single crystal neutron diffraction data and those measured on powder samples which arises from a difference in the respective transition temperatures, there is a good qualitative agreement between these two sets of data. 
 
\begin{figure}
\includegraphics[width=0.99\columnwidth]{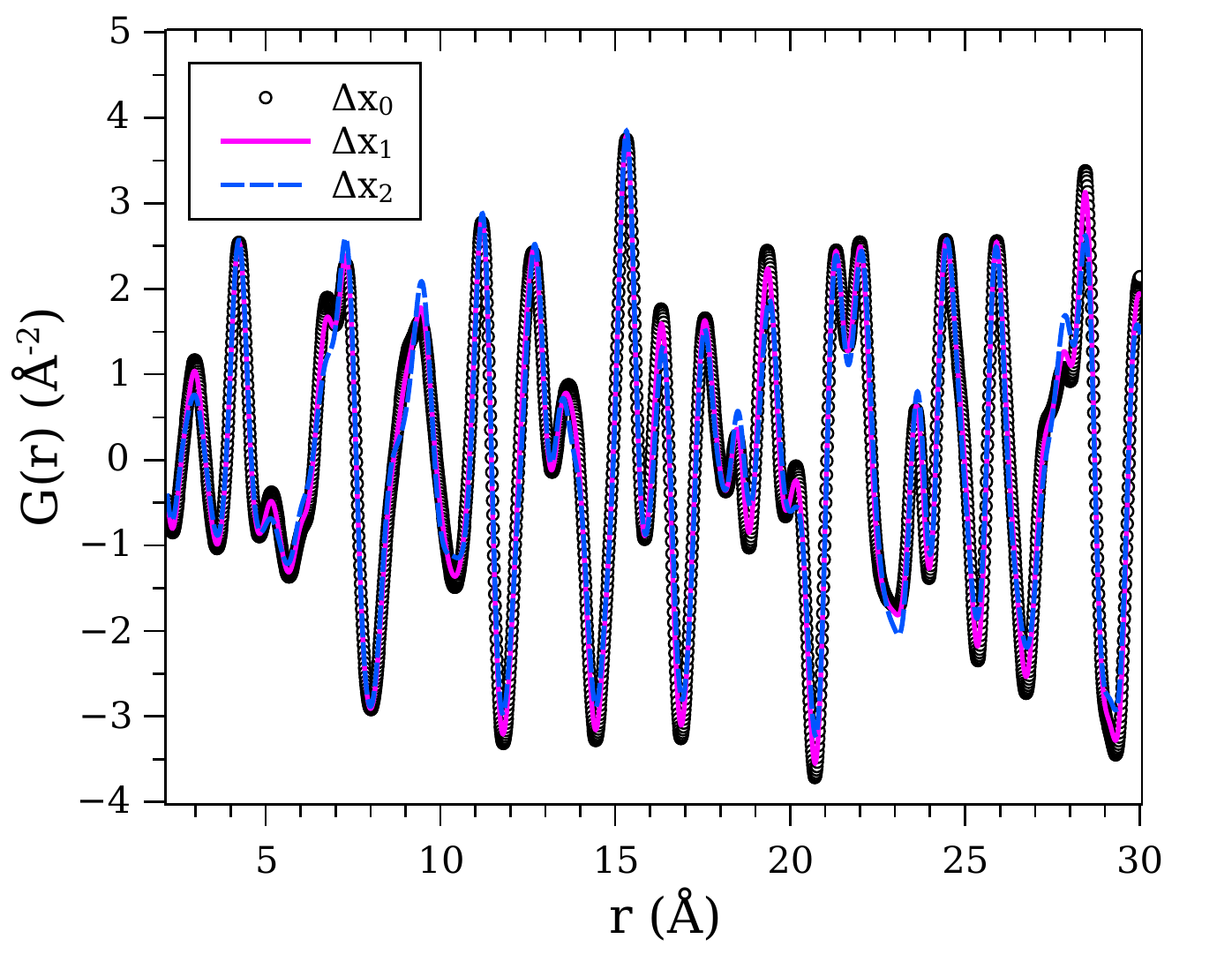}
\caption {(Color online) Pair distribution function $G(r)$ calculated for the cubic GeTe phase with the static distortions $\Delta x_0 = 0$, $\Delta x_1 = 7.94 \times 10^{-3}$, and $\Delta x_2 = 14.56 \times 10^{-3}$. Simulation are performed at the experimental $r$-range and $Q_{max}=15$~\AA$^{-1}$ \cite{matsunaga11}.}
 \label{rdf}
\end{figure}

 Such a behavior as revealed by the Ge--Te bond lenghts in both single crystal and powder diffraction studies is however claimed \cite{fons10} to be the case only in the so-called average structure. The PDF analysis of the total diffraction data on GeTe shows that the high temperature phase exhibits two distinct bond lengths which hardly change across the phase transition  \cite{matsunaga11}. The local scale structural distortions, evidenced by unequal Ge--Te bond distances above $T_c$, suggested that the transition in GeTe could be of the order-disorder type \cite{fons10,matsunaga11}. To obtain additional information about the local structure of GeTe above $T_c$, we have simulated the PDF spectra of its high temperature phase with the static structural lattice distortions generated according to the triply degenerate unstable soft-phonon mode of $\Gamma_{4}^{-}$ symmetry \cite{wdowik14}. This mode, while frozen, leads to relative displacements of the Ge and Te sublattices along the cubic cell diagonal. Results of these calculations for perfectly ordered ($\Delta x_0 = 0$) and disordered ($\Delta x_1 = 7.94 \times 10^{-3}$ and $\Delta x_2 = 14.56 \times 10^{-3}$) high temperature cubic phases of GeTe are shown in Fig.~\ref{rdf}. The distortions  $\Delta x_1$ and $\Delta x_2$ correspond respectively to the Ge--Te bond lengths of ($s_1=2.91$~\AA, $l_1=3.10$~\AA) and ($s_2=2.84$~\AA, $l_2=3.19$~\AA).  Here we note that $\Delta x_2$ also reflects the Ge--Te bond lengths determined above $T_c$ by using the experimental PDF data \cite{matsunaga11}. We observe that the simulated spectra of cubic GeTe with and without static distortions are essentially identical. Moreover, they are closely related to the experimental PDF data reported by Matsunaga \textit{et al.} \cite{matsunaga11}.  This finding enables us to suggest that the PDF results can hardly provide a definite answer about the displacive or order-disorder type of the phase transition in GeTe compound since they probe the average static lattice distortions but not a dynamical nature of this transition connected with the phonon dynamics \cite{wdowik14}. 

\begin{figure}[h!]
\includegraphics[width=0.98\columnwidth]{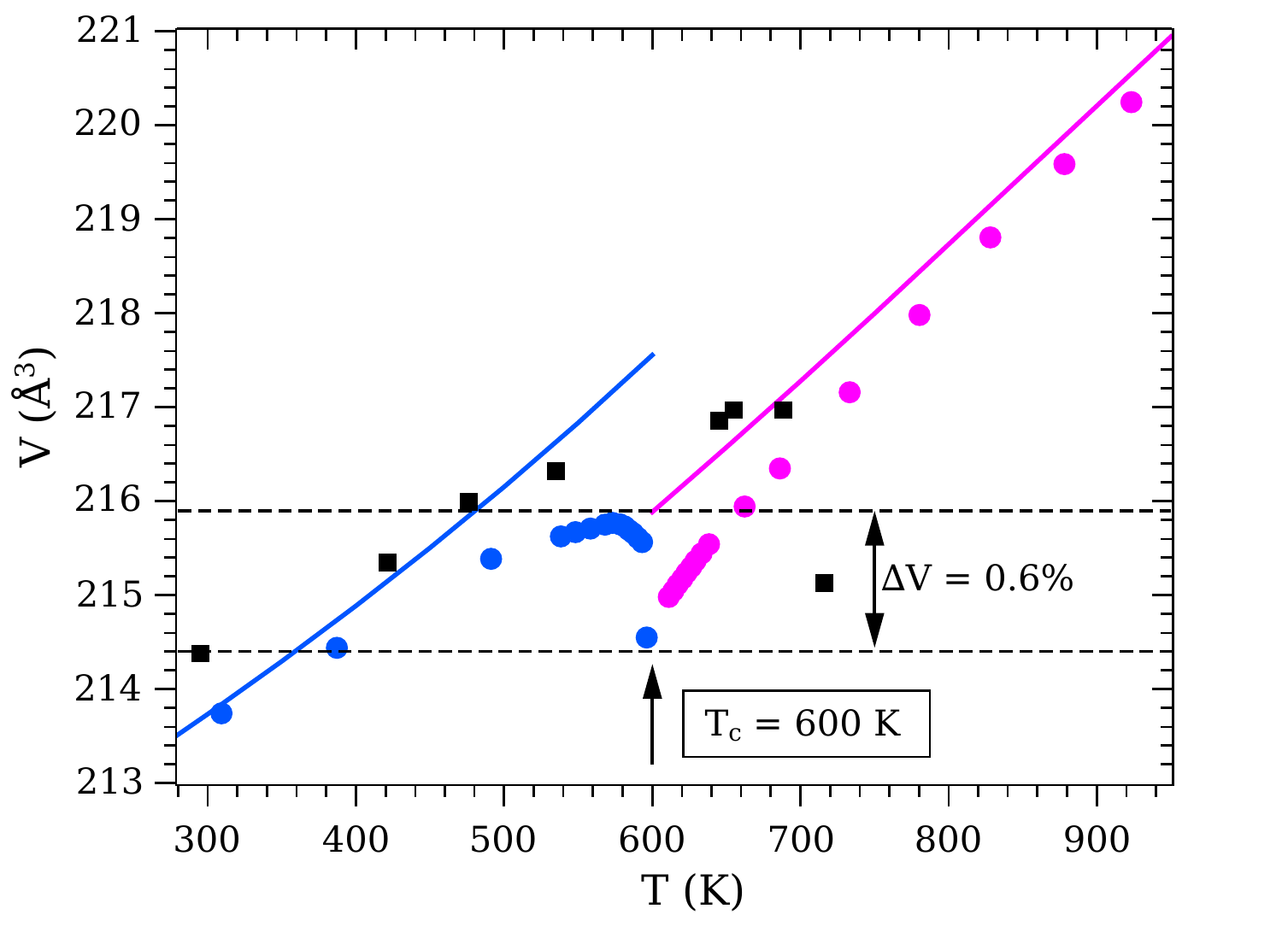}
\caption {(Color online) Temperature dependence of the GeTe pseudo-cubic unit cell volume ($V$). The present neutron powder diffraction results (circles) are compared to the single crystal neutron diffraction data obtained by Chattopadhyay \textit{et al.} \cite{chattopadhyay87}. Solid lines denote results of the DFT-QHA calculations for the low and high temperature phases of GeTe.}
 \label{volume}
\end{figure}

Figure \ref{volume} indicates that both powder and single crystal samples of GeTe exhibit  temperature-induced volume reduction $\Delta V$ at the ferroelectric rhombohedral-to-cubic phase transition. The unit cell volume decreases by $\Delta V\approx0.6$\% at $T_c=600$~K, as indicated by our NPD data.
The volume anomaly at the ferroelectric structural transition in GeTe can be considered as an electrostriction effect in analogy with magnetostriction effects at the magnetic ordering. Also, the volume contraction at $T_c$ has been interpreted by Chattopadhyay \textit{et al.} \cite{chattopadhyay87} as the extra volume due to the presence of lone pairs in the rhombohedral phase and the absence of this excess volume in the high temperature cubic phase. Results of the DFT and QHA calculations performed for both the low temperature rhombohedral and high temperature cubic structures of GeTe support to some extent the present experimental data. The agreement between calculations and experiment remains reasonable, except the close vicinity of $T_c$, where the QHA approach is unable to reproduce the volume reduction at the onset of the phase transformation. Our calculations indicate that the volume of the rhombohedral phase is higher than the volume of the cubic phase. Indeed, in the rhombohedral structure the Ge and Te atoms are displaced from each other with respect to their position in the cubic structure. The relative shift of the Ge and Te sublattices along the body cell diagonal ($\Delta x$) is accompanied by angular distortion of the GeTe lattice ($\Delta \alpha$) as well. Both distortions result in the larger volume of the rhombohedral structure in comparison with the rocksalt one. These distortions are dynamical in their origin as they are driven by the soft phonon mode $\Gamma_4^{-}$, as shown and discussed in our recent paper \cite{wdowik14}. The DFT calculations show that the rhombohedrally distorted GeTe is more energetically stable than the undistorted cubic one. At the ground state the difference in their Helmholtz free energies amounts to 26~meV per formula unit, i.e., it lies in the range of thermal excitations. The present experiments also show that each of GeTe phases expands upon heating outside the temperature range where the phase transition occurs. The volume thermal expansion coefficient of the rhombohedral GeTe equals $4.59\times10^{-5}$~K$^{-1}$ at 300~K, whereas it amounts to $7.67\times10^{-5}$~K$^{-1}$ at 650~K for the cubic GeTe.       
 
\begin{figure}[h!]
\includegraphics[width=0.98\columnwidth]{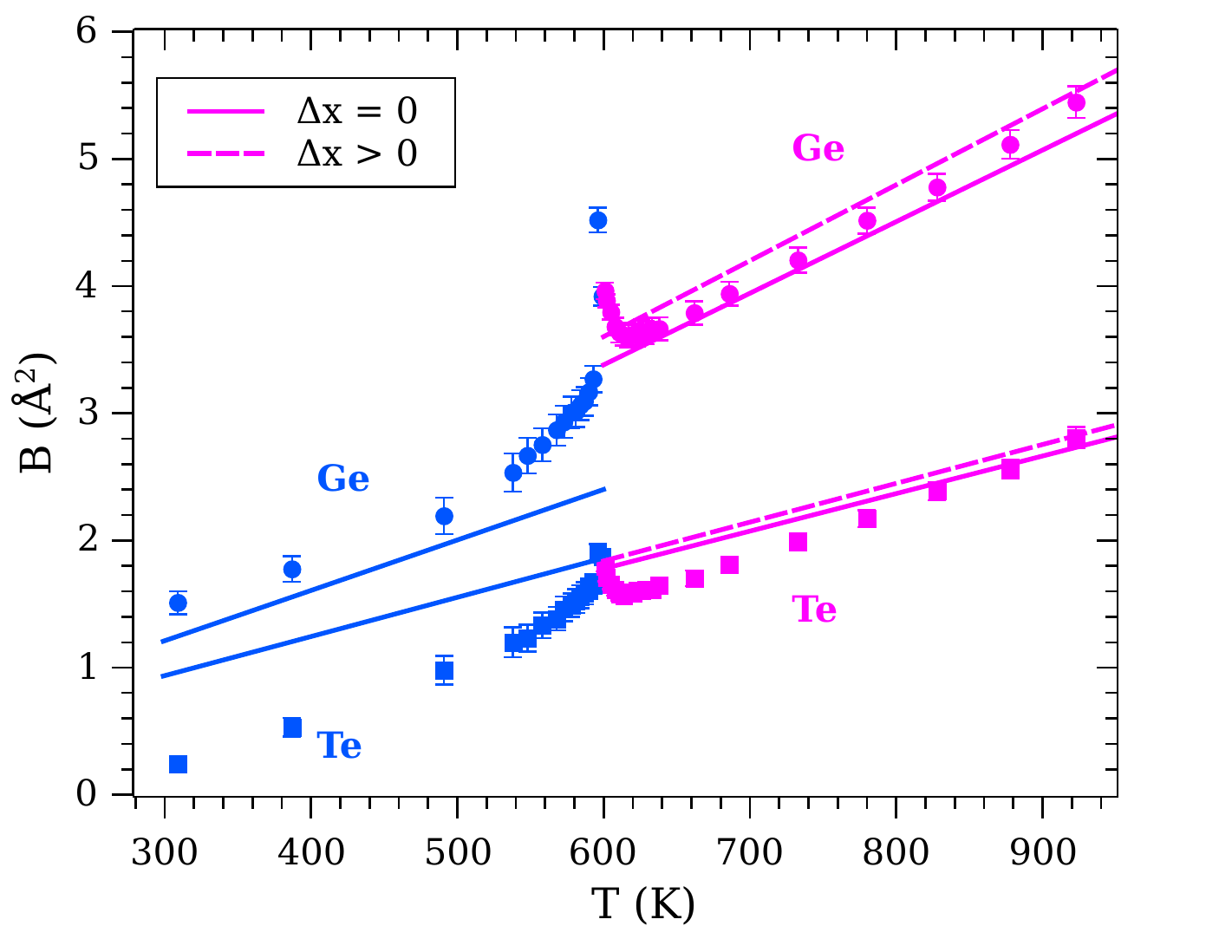}
\caption {(Color online) Experimental (symbols) and theoretical (lines) isotropic temperature factors $B$ determined for Ge (circles) and Te (squares) atoms in the rhombohedral and cubic phases of GeTe.  The solid lines denote calculated $B$ factors for the rhombohedral and ideal cubic GeTe structures. Dashed lines correspond to calculated $B$ factors for the cubic GeTe structure with static distortion $\Delta x = 14.56\times10^{-3}$ ($s=2.84$~\AA, $l=3.19$~\AA). Calculations are performed within the harmonic approximation. Theoretical isotropic temperature factors are evaluated as $B=8\pi^2\left<U\right>$, where $\left<U\right>$ is the trace of $U_{ij}$ tensor \cite{wallace72}.}
\label{Bfac} 
\end{figure}

The phase-change material GeTe is not a unique system undergoing temperature-induced volume collapse at the phase transition. The  well-known example is the volume contraction of ice at its melting temperature.  This phenomenon remains, however, unexplained quantitatively so far.  The temperature-induced volume reduction is also found to exist in solid-to-solid phase transitions in diverse condensed matter systems. Even earlier-known industrially important canonical ferroelectric material like BaTiO$_3$ shows such volume decrease at its ferroelectric phase transition at 393~K \cite{shirane52}. One notable example is the orbital order-disorder transition or orbital melting in the strongly correlated electron system LaMnO$_3$ \cite{chatterji03} -- the recognized parent compound of colossal magnetoresistive manganites that lure a host of condensed matter scientists for spintronics and other device applications. Reduction of the LaMnO$_3$ volume at the orbital order-disorder transition is assisted by the considerable change in the atomic mean-squared vibrational amplitudes \cite{chatterji03}. Very similar effect is observed in the present studies on GeTe too. The isotropic temperature factors $B_{\mathrm{Ge}}$ and $B_{\mathrm{Te}}$ depicted in Fig.~\ref{Bfac} increase with increasing temperature and show $\lambda$-type behavior in the close vicinity of the rhombohedral-to-cubic phase transition. The $B_{\mathrm{Ge}}$ and $B_{\mathrm{Te}}$ measured for single crystals of GeTe \cite{chattopadhyay87} are much more scattered compared to the present data. They are also limited to temperatures not exceeding $T_c$ and therefore they do not reveal such a characteristic change at $T_c$ as  $B_{\mathrm{Ge}}$ and $B_{\mathrm{Te}}$ in our experiments.

It is well know that the atomic temperature factors remain sensitive to the immediate atomic environment. Experimental atomic thermal displacements contain both static and dynamic effects. The dynamic contribution to the atomic thermal displacements is gained from the present DFT calculations. The $U_{ij}$ tensor for both Ge and Te atoms in the $R3m$ structure contains two independent components, namely $U_{xx} = U_{yy}$ and $U_{zz}$ which represent the atomic thermal vibrations perpendicular and parallel to the three-fold rotational axis, respectively (cf. Fig.~\ref{structure}). The site symmetries of the Ge and Te atoms in the $Fm\bar{3}m$ structure constrain their mean-squared displacements to be isotropic and hence described by a single parameter. Our calculations show that the thermal motions of Ge and Te atoms in the rhombohedral GeTe are almost isotropic as $U_{xx}^{\mathrm{Ge}}\approx U_{zz}^{\mathrm{Ge}}$ and $U_{xx}^{\mathrm{Te}}\approx U_{zz}^{\mathrm{Te}}$. Similar dependences were also observed in the neutron diffraction study on single crystals of GeTe \cite{chattopadhyay87}. The nearly isotropic $U_{ij}$ tensor arises from very small difference between the atomic on-site force constants along and perpendicular to the three-fold rotational axis in the rhombohedral GeTe. This negligible difference remains in accordance with the structural features of the rhombohedral phase for which the structural distortions of the Peierls-type are indeed small. The mean-squared vibrational amplitudes are obviously different for the Ge and Te atoms due to the difference in their masses. Also, they exhibit typical increase with increasing temperature according to the applied harmonic approximation with the $U_{ij}$ growing their values over two times between 300 and 600~K. The harmonic approach is, however, unable to describe properly the anomalous behavior of $U_{ij}$ in the close vicinity of $T_c$. Nevertheless, it occurred to be helpful in explaining the behavior of atomic thermal vibrations in the cubic phase of GeTe. We notice that the slope of $B_{\mathrm{Te}}$ practically does not change it course with increasing temperature, except the tiny range of temperatures around $T_c=600$~K when it shows small discontinuity. There is, however, considerable jump in the thermal vibrations of the Ge atoms in the cubic GeTe phase as well as visible change in the slope of $B_{\mathrm{Ge}}$ above $T_c$. The results of our calculations show the the on-site force constants of Ge atoms in the cubic GeTe are almost two times lower compared to the respective on-site force constants in the rhombohedral GeTe structure, i.e., the Ge atoms in the cubic phase are more loosely bound inside the lattice which in turn allows for the larger values of the amplitudes of their mean-squared displacements. The meaningful decrease in the on-site force constants of the Ge atoms while going from the rhombohedral to cubic GeTe accounts for the increase in $B_{\mathrm{Ge}}$ slope above $T_c=600$~K. It is interesting to note that the cubic phase of GeTe with incorporated static lattice distortions of Peierls-type $\Delta x$ shows higher $B$-factors in comparison with the respective factors for the ideal cubic structure ($\Delta x = 0$). In addition, the effect of $\Delta x>0$ is much more pronounced for $B_{\mathrm{Ge}}$ than for $B_{\mathrm{Te}}$.  Our calculations indicate that progressive increase of $\Delta x$ results in the gradual growth and small changes in slopes of $B$-factors due to modified force constants which decrease with decreasing $\Delta x$. In principle, this observation allows to distinguish the high temperature GeTe structure with persisting static lattice distortions (retained distinct short and long Ge--Te bonds) above the phase transition from that with the local distortions vanished (unique Ge--Te bond length).
  
\section{Summary and conclusions}
Chattopadhyay \textit{et al.} \cite{chattopadhyay87} carried out high temperature single crystal neutron diffraction investigations on GeTe almost three decades ago when the powder neutron diffraction technique was not very mature. The present NPD investigations on GeTe have been carried out on a modern high resolution powder diffractometer at a spallation neutron source and have benefited from the enormous progress made during recent years in the neutron powder diffraction technique. The high temperature single crystal neutron diffraction is very time consuming and requires temperature stability for long periods whereas neutron powder diffraction is relatively free from these constraints. The extinction effects can be very large in single crystal diffraction whereas it is often practically negligible in powder diffraction. Also, the NPD technique appears much attractive in describing the phase transitions in relatively simple structures like GeTe due to its less sensitivity to the crystal domain structure which significantly complicates both the data collection and the data treatment. 

The results of our NPD measurements on GeTe are based on Rietveld refinement involving only  Bragg intensities. An information contained in the background diffuse scattering from the sample is not taken into account, and hence our results are related to the average structure, but not to the local structure or dynamics. Although the current experimental results could not unambiguously resolve  controversy about the nature of the phase transition in GeTe, viz. whether this phase transition is of displace or order-disorder type, they allowed for more accurate  determination of the temperature dependences of the lattice and structural parameters of GeTe compared to the previous studies \cite{chattopadhyay87}. The measured variation of the GeTe volume over a wide temperature range is however a robust result of Bragg diffraction and the volume anomalous behavior at the phase transition is the most important result. The temperature variation of the structural parameters across the rhombohedral-to-cubic phase transition provides us microscopic mechanism behind the volume discontinuity and enables us to remove the veil of mystery around this transition.

It is interesting to note that the interpretation of the local probes results \cite{fons10,matsunaga11} viz. the PDF analysis of the total diffraction intensities or EXAFS, give an impression that apart from a small linear thermal expansion nothing happens during the phase transition, namely the short and long Ge--Te bond lengths do not change within experimental errors in the whole wide temperature range investigated (300--800~K). One also wonders whether a hypothetical static model compatible with the bond distances obtained by the local probes would reproduce the robust result of volume contraction at the rhombohedral-to-cubic phase transition in GeTe. Perhaps some kind of a dynamical model, though definitely difficult to construct, might reproduce the volume decrease. Hence, an answer to the question whether transition is of displacive or order-disorder type lies in the dynamics of phase transition and the role of soft modes, as has been recently shown by the DFT and phonon calculations \cite{wdowik14}. The conventional PDF analysis of the total scattering data does not analyze the energy and therefore it lacks an information about dynamics of the phase transition. The high temperature phase contains the dynamics of broken symmetry phase and the PDF local probe just sees the snap-shot of low frequency soft phonon mode that happened to enter into the window of local probe. The two distinct Ge--Te bond lengths seen by the local probes at the high temperature phase are just dynamical signature of the low temperature phase still persisting at temperatures exceeding $T_c$. Since the PDF measures instantaneous structure and cannot distinguish between static and dynamic correlations, interpretation of the PDF results based solely on the static bond distances could be verified.

\section{Acknowledgments}
The research conducted at SNS was sponsored by the Scientific User Facilities Division, Office of Basic Energy Sciences, US Department of Energy. Interdisciplinary Center for Mathematical and Computational Modeling (ICM), Warsaw University, Poland and the IT4Innovations National Supercomputing Center, VSB-Technical University,
Ostrava, Czech Republic are acknowledged for providing the computer facilities under Grants No.~G28-12 and Reg.~No.~CZ.1.05/1.1.00/02.0070.

\end{document}